\documentclass[prb,aps,amsmath,amssymb,reprint]{revtex4-1}
\usepackage{graphicx}% Include figure files
\usepackage{dcolumn}% Align table columns on decimal point\bibliographystyle{model1a-num-names}
\usepackage{bm}% bold math
\newcommand{\p}{$\%$}
\newcommand{\pat}{{${~at.}$}\%}

\newcommand{\pn}{$\mathrm{R{_{N_2}}}$}

\newcommand{\tfn}{$\mathrm{Fe_{4}N}$}
\newcommand{\tcn}{$\mathrm{Co_{4}N}$}

\newcommand{\tmn}{$\mathrm{M_{4}N}$}

\newcommand{\Ts}{$\mathrm{T_{s}}$}

\begin{document}

\title{Finding pathways for stoichiometric \tcn~thin films}
\author {Nidhi Pandey$^1$, Mukul Gupta$^{1*}$, Rachana Gupta$^2$, Zaineb Hussain$^1$, V. R. Reddy$^1$, D. M. Phase$^1$ and Jochen Stahn$^3$}
\address{$^1$UGC-DAE Consortium for Scientific Research, University
Campus, Khandwa Road, Indore 452 001, India}
\address{$^2$Institute of Engineering and Technology DAVV, Khandwa Road, Indore 452 017, India}
\address{$^3$Laboratory for Neutron Scattering and Imaging, Paul Scherrer Institut, CH-5232 Villigen PSI, Switzerland}
\author {}
\address{$^*$ Corresponding author email: mgupta@csr.res.in}

\begin{abstract}

In this work, we studied the pathways for formation of
stoichiometric \tcn~thin films. Polycrystalline and epitaxial
\tcn~films were prepared using reactive direct current magnetron
(dcMS) sputtering technique. A systematic variation in the
substrate temperature (\Ts) during the dcMS process reveals that
the lattice parameter (LP) decreases as \Ts~increases. We found
that nearly stoichiometric \tcn~films can be obtained when \Ts~=
300\,K. However, they emerge from the transient state of Co target
($\phi$3\,inch). By reducing the target size to $\phi$1\,inch, now
the \tcn~phase formation takes place from the metallic state of Co
target. In this case, LP of \tcn~film comes out to be
$\sim$99\p~of the value expected for \tcn. This is the largest
value of LP found so far for \tcn. The pathways achieved for
formation of polycrystalline \tcn~were adopted to grow an
epitaxial \tcn~film, which shows four fold magnetic anisotropy in
magneto-optic Kerr effect measurements. Detailed characterization
using secondary ion mass spectroscopy indicates that N diffuses
out when \Ts~is raised even to 400\,K. Measurement of electronic
structure using x-ray photoelectron spectroscopy and x-ray
absorption spectroscopy further confirms it. Magnetization
measurements using bulk magnetization and polarized neutron
reflectivity show that the saturation magnetization of
stoichiometric \tcn~film is even larger than pure Co. Since all
our measurements indicated that N could be diffusing out, when
\tcn~films are grown at high \Ts, we did actual N self-diffusion
measurements in a CoN sample and found that N self-diffusion was
indeed substantially higher. The outcome of this work clearly
shows that the \tcn~films grown prior to this work were always N
deficient and the pathways for formation of a stoichiometric
\tcn~have been achieved.

\end{abstract}

\date{\today}
%\begin{keyword}
%epitaxial Co4N thin films, tetra transition metal nitrides, reactive sputtering, N self-diffusion, high magnetic moment

%\end{keyword}

\maketitle
\section{Introduction}
\label{intro}

\begin{figure}
\centering
\includegraphics[scale=0.45]{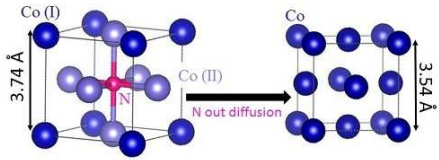}
\caption{\label{fig:fcc} (Color online) Schematic of conversion of
\tcn~to fcc Co by N out diffusion.}
\end{figure}

Tetra transition metal nitrides of magnetic metals (M$_4$N; M =
Fe, Co and Ni) are a class of compounds that have an
anti-perovskite type fcc structure in which N atom resides at
one-quarter of the octahedral interstices in an ordered manner.
The insertion of N in the fcc metal lattice results in two
inequivalent metal sites: one occupying the corner (M I) and
others at face center (M II) positions of the cube as shown in
fig.~\ref{fig:fcc} (a). Here M II sites get hybridize with N atom
but M I remain isolated from N. This results in localized
magnetism at M I sites whereas itinerant magnetism at M II sites.
The insertion of N causes an expansion in the lattice (compared to
fcc host metal lattice)~\cite{PRB07_Matar, JMMM10_Imai,
JAC14_Imai, Coey.JMMM.1999, JMMM99_P_Mohn_Matar, JPCM16_Markus}
and leads to very interesting magnetic properties such as higher
saturation magnetization (Ms) due to a magneto volume effect.

The spin polarization ratio (SPR) of \tmn~is higher than their
host metals.~\cite{PRB06_Kokado_Fe4N_SPR, JMMM10_Imai} Polarized
electronic band structures of these \tmn~compounds were calculated
using the full-potential method.~\cite{JMMM11_Takahashi_Matar_SPR,
JMMM10_Imai} Among all \tmn, \tcn~is predicted to be the most
effective material for spin-polarization of conducting electrons
as its spin polarization ratios at the Fermi level reaches to
about 90\p.~\cite{JMMM11_Takahashi_Matar_SPR, JMMM10_Imai} In
addition, the insertion of N reduces the corrosion of metal and
therefore M$_4$N compounds can be considered as a superior
alternative to their pure metals in magnetic devices
applications.~\cite{JMMM91_Fe4N_Sakuma, PRB93_Coehoorn_Fe4N,
PRB06_Kokado_Fe4N_SPR}

Larger Ms has been predicted theoretically for all M$_4$N
compounds and observed experimentally for \tfn. Whereas in case of
\tcn~the value of Ms has always been found to be lower than the
pure Co. In this connection, it may be noted that the theoretical
value of lattice parameter (LP) for \tcn~is
3.735\,{\AA}~\cite{PRB07_Matar} but experimentally, it has been
found anywhere between 3.59 to 3.52\,{\AA}.~\cite{JMS87_Oda,
JAC15_Silva, 2011_Co4N_K_Ito, JAP14_Ito, TSF14_Silva,
Wang:CoN:TSF:09, MSEB08_Jia, Vac01_Asahara, JVSTA04_Fang} Since
the LP of fcc Co is about 3.54\,{\AA},~\cite{PRB07_Matar} it
appears that \tcn~thin films deposited so far might have mistaken
for fcc Co. This can be envisaged by closely following the
substrate temperature (\Ts) used for deposition of \tcn~phase. We
find that \Ts~between 435 to 725\,K have been used and a
correlation can be found between \Ts~and LP. Oda $et$. $al$. first
attempted to prepare \tcn~films at \Ts~= 435\,K and got LP =
3.59\,{\AA},~\cite{JMS87_Oda} at \Ts~= 525\,K Silva $et$. $al$.
found LP = 3.54\,{\AA}.~\cite{JAC15_Silva} When the \Ts~was even
raised to 700\,K and beyond, the LP remains at 3.54\,{\AA} a value
expected for fcc Co (not for \tcn).~\cite{2011_Co4N_K_Ito,
JAP14_Ito} It apparently looks that as \Ts~is raised N diffuses
out from the system, leaving behind fcc Co as shown in
fig.~\ref{fig:fcc}. Theoretically, it is expected that as
N\pat~decreases in \tcn~the LP should also decrease
monotonically.~\cite{PRB07_Matar, JAC14_Lourenco} Therefore, the
LP of \tcn~can be used for estimating N deficiency in \tcn. In our
recent works we studied phase formations in Co-N system and found
that at \Ts~= 525\,K, fcc Co is formed, irrespective of the amount
of N$_2$ gas used during reactive sputtering,~\cite{JAC16_NPandey}
however as the \Ts~goes down to 300\,K, variety of Co-N phases
emerge including \tcn.~\cite{AIP15_RG}

In absence of the phase diagram of Co-N system, it is difficult to
assess the role of \Ts~on the phase formation of \tcn~thin films.
Since the formation of \tfn~thin films takes place typically at
\Ts~= 675\,K, similar high \Ts~have been used for deposition of
\tcn~films as well. The enthalpy of formation
($\Delta$H$^{\circ}_f$), is an important parameter that affects
formation of any compound. For \tfn, $\Delta$H$^{\circ}_f$
$\sim$-12.17$\pm$20.72\,kJ/mol~\cite{Tessier_SSS00} but for
\tcn~it is significantly higher at about
0$\pm$2.896\,kJ/mol.~\cite{JAC14_Imai} This may have consequences
on the \tcn~phase formation that are addressed in this work.

First of all, we studied the effect of \Ts~on the phase formation
of \tcn~thin films. In section~\ref{smc}, crystal structure,
composition and morphology were studied and it was found that as
\Ts~reduces the LP increases and its maximum value was found when
\Ts~= 300\,K. Here, it can be anticipated that N diffuses out when
\Ts~is raised. This assumption was further confirmed by measuring
the electronic structure shown in section~\ref{elec}. Here both
XPS and XAS measurements confirm that sample becomes highly N
deficient even at \Ts~= 400\,K. Magnetization measurements
presented in section~\ref{mag} clearly show a loss in saturation
magnetization with LP in agreement with theoretical predications
for \tcn. In section~\ref{tp}, we show that by reducing the target
size, the formation of \tcn~takes place from the `metallic state'
of target and then fully stoichiometric \tcn~phase can be
realized. Under optimized conditions, epitaxial \tcn~was grown and
it exhibits a four fold magnetic anisotropy expected for cubic
symmetry as discussed in section~\ref{epi}. From all these
measurements, it becomes obvious that N diffusion should be much
larger in the Co-N system as compared to similar systems (e.g.
Fe-N). Since N self-diffusion in Co-N was not measured before, we
measured N self-diffusion in CoN and compared it to FeN as shown
in section~\ref{n_diff}. Overall, the results presented here aims
to find the pathways for a stoichiometric \tcn. After achieving
this, detailed analysis of structural, electronic and magnetic
properties is presented and discussed in this work.

\section{EXPERIMENT} \label{expe}

A series of Co-N thin films were prepared on different substrates
(SiO$_2$, Si and LaAlO$_3$ (100)) using direct current magnetron
sputtering (dcMS) system (Orion-8, AJA Int. Inc.). Co target
(99.95\p) ($\phi$ = 3\,inch) were sputtered using a mixture of
argon (99.9995\p) and nitrogen (99.9995\p) gases flowing through
two different mass flow controllers. The partial gas flow of
nitrogen ({\pn} =
p$_{\mathrm{N}_2}$/(p$_{\mathrm{Ar}}$+p$_{\mathrm{N}_2}$), where
p$_{\mathrm{Ar}}$ and p$_{\mathrm{N}_2}$ are gas flow of Ar and
N$_2$ gases, respectively) was kept at 20\p, a value favoring in
growth of \tcn~phase.~\cite{JAC16_NPandey, JMMM17_NPandey,
AIP15_RG} A base pressure of 2$\times$10$^{-7}$\,Torr~was achieved
before deposition and the working pressure was maintained at
3\,mTorr~during deposition. Keeping all other parameters constant,
substrate temperature (\Ts) was kept at 300 (no intentional
heating), 325, 350, 375 and 400\,K. The thicknesses of the films
were ranging from 1200 to 1400\AA. In addition, another set of
Co-N thin films were also grown by $\phi$ = 1\,inch Co target
(99.95\p~purity) at different values of \pn~= 0, 5, 10, 12, 20,
25, 30, 40 and 100\p~where \Ts~was kept at 300\,K~(all other
deposition parameter were kept same as above mentioned).

Deposited samples were characterized for their long range ordering
and phase formation by x-ray diffraction (XRD) using a standard
x-ray diffractometer (Bruker D8 Advance) having CuK$\alpha$ x-ray
source. To get the information about the atomic concentration of
nitrogen in deposited samples, secondary ion mass spectroscopy
(SIMS) depth profiling measurements were carried out using a Hiden
Analytical SIMS workstation. A primary O$_2$$^+$ ions source was
used for sputtering with an energy of 4\,keV and beam current of
200\,nA. SIMS measurements were performed in a UHV chamber with a
base pressure of the order of 5$\times$10$^{-10}$\,Torr~while
during measurements the chamber pressure was
1$\times$10$^{-7}$\,Torr. The sputtered secondary ions were
detected using a quadrupole mass analyzer. Morphology of the
deposited samples were studied by atomic force microscopy(AFM) in
non contact mode. Electrical resistivity was measured in a four
probe mode at room temperature. X-ray photoelectron spectroscopy
(XPS) measurements were carried out using Al-K$\alpha$ lab source.
Prior to XPS measurement, samples were sputtered with an ion
energy of 1\,kV and current of 4\,$\mu$A. To investigate the local
and electronic structure, x-ray absorption near edge spectroscopy
(XANES) was performed in the total electron yield (TEY) mode at
BL-01~\cite{XAS_beamline} at the Indus-2 synchrotron radiation
source at RRCAT, Indore. The XANES measurements were carried out
in a UHV chamber with a base pressure of
2$\times$10$^{-10}$\,Torr. To avoid surface contaminations,
samples were cleaned $in-situ$ using a Ar$^+$ source with an
energy of 5\,keV~kept incident at an angle of 45$^{\circ}$. The
magnetization measurements were carried out at room temperature
using a Quantum Design SQUID-VSM (S-VSM) magnetometer. Polarized
neutron reflectivity (PNR) measurements were performed at AMOR
beamline, SINQ, PSI Switzerland in time of flight mode. During PNR
measurements, to saturate the sample magnetically, a magnetic
field of 0.5\,T was applied parallel to the sample's surface.
Reciprocal space mapping(RSM) and Phi scans were performed with
high resolution X-ray diffractometer (Bruker D8 Discover).
Magnetic anisotropy was studied in longitudinal mode using magneto
optical-Kerr effect (MOKE) and Kerr microscopy (Evico Magnetics)
equipment.

\section{Results and Discussion} \label{res}
\subsection{\textbf{Structure, morphology and  composition of \tcn~films}}
\label{smc}

\begin{figure} \center
\vspace{-1mm}
\includegraphics [width=80mm,height=85mm] {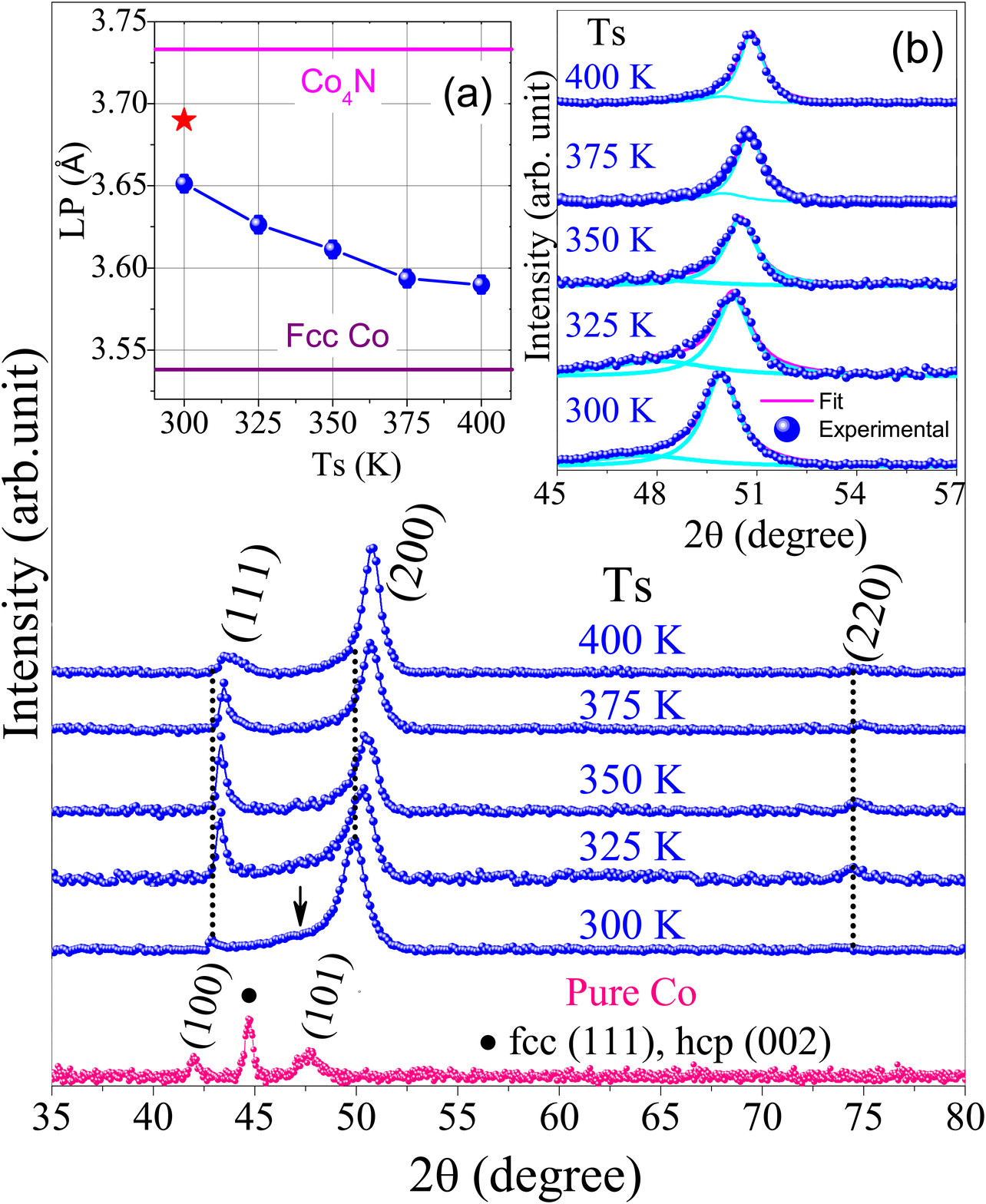}
\caption{\label{fig:xrd} (Color online) XRD patterns of Co-N thin
films deposited at different \Ts~from 300 to 400\,K. Inset showing
the variation of LP with respect to \Ts~(a) and fitted XRD curves
along (200) reflection (b). The error bars in estimation of LP are
typically the size of symbols used here.} \vspace{-1mm}
\end{figure}

Fig.~\ref{fig:xrd} shows the XRD patterns of Co-N thin films
deposited at \Ts~= 300, 325, 350, 375 and 400\,K. For reference,
XRD pattern of a Co thin film of similar thickness is also
included here. In the Co thin film, we can observe peaks at 41.9
and 47.5$^{\circ}$ corresponding to (100) and (101) planes,
respectively for hcp structure (JCPDS
05–0727).~\cite{1999_Co_H_Zhang} In addition, the peak at
44.7$^{\circ}$ coincides with (111) plane of fcc and (002) plane
hcp Co. Generally, Co stabilizes in the hcp structure under
ambient conditions and show an allotropic transition from hcp to
fcc at high temperature (above 690\,K). But, the occurrence of hcp
and fcc biphase usually takes place in thin films and
nanoparticles.~\cite{1999_Co_H_Zhang, MSE01_RAM_Co_nano}

\begin{figure} \center
\vspace{-1mm}
\includegraphics [width=83mm,height=68mm] {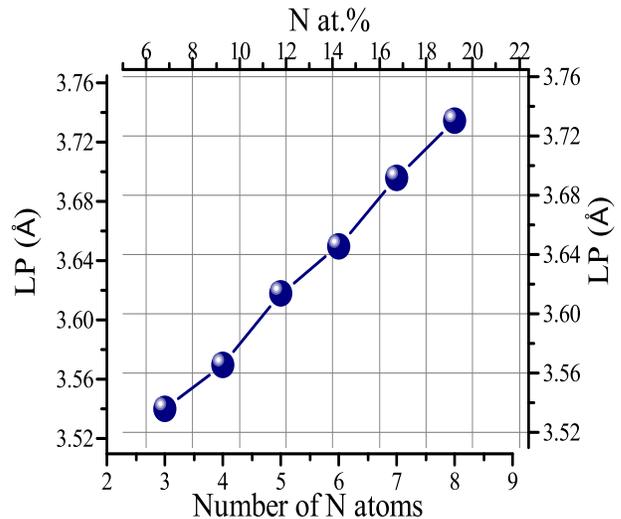}
\caption{\label{fig:lp_n} (Color online) Theoretical correlation
of LP with N\pat~and no. of N atom in a supercell (having 8 unit
cell) of \tcn~lattice.} \vspace{-1mm}
\end{figure}

The XRD patterns of samples prepared using partial nitrogen gas
flow, show prominent peaks around 43, 50 and 74$^{\circ}$ and they
can be respectively assigned to (111), (200) and (220) planes of
fcc \tcn~structure.~\cite{JMS87_Oda} We can clearly see as the
\Ts~is raised, peaks gradually shift towards higher 2$\theta$
values. This immediately indicates that the LP (obtained from
(200) peak) is decreasing with an increase in \Ts~as shown in the
inset (a) of fig.~\ref{fig:xrd}. Such lowering of LP may be due to
N deficiency in \tcn. In recent theoretical studies, it was shown
that the LP is directly correlated with the N\pat~in \tcn. It was
found that as the amount of N decreases (considering a supercell
of 8 unit cells of \tcn) the LP also reduces~\cite{PRB07_Matar,
JAC14_Lourenco} as shown in fig.~\ref{fig:lp_n} and therefore LP
may be used to estimate N concentration in \tcn. Table.~\ref{tab}
compares values of LP and N\pat~measured experientially using the
formalism of Mater $et$. $al$.~\cite{PRB07_Matar} From here, it is
clear that N\pat~and LPs get reduced as \Ts~increased, which
elucidate that a compression in the \tcn~lattice takes place
probably due to N out diffusion from the lattice at higher \Ts.

As we find that the LP increases by decreasing \Ts, at the lowest
\Ts~= 300\,K, obtained value of LP = 3.65\,{\AA}. This values is
about 3\p~smaller than the theoretical value (3.735\,{\AA}). To
further enhance it, we have addressed the issues related with
target poisoning in reactive sputtering and they are discussed in
detail in section~\ref{tp}. We find that when the \tcn~phase
emerges out of the metallic state of the target, LP can be
enhanced further to 3.7\,{\AA}. This is the highest value of LP of
\tcn~found so far as shown by an asterisk ($\ast$) in the inset
(a) of fig.~\ref{fig:xrd}.

In addition, we found that the XRD peak shape of Co-N samples
shows an asymmetry, especially for those with \Ts$\leq$350\,K.
This asymmetry is marked by an arrow ($\downarrow$) for \Ts~=
300\,K sample and can be fitted if we consider (101) reflection of
hcp Co$_3$N$_{1-x}$. It indicates that \tcn~formed with small
amount of Co$_3$N$_{1-x}$ phase. At higher \Ts~($\geq$350\,K) this
asymmetry disappears indicating that hcp phase transforms into
fcc.

\begin{table}\center
\caption{\label{tab} Relationship between LP and N\pat~for
\tcn~determined experimentally from XRD and SIMS measurements.}

\begin{tabular}{ccccc}\hline
\Ts&&LP&N (XRD)&N (SIMS)\\
(\,K)&&(\AA)&(\pat)&(\pat)\\\hline\hline

300&&3.65&15&15$\pm$2\\
325&&3.63&13&-\\
350&&3.61&12&12$\pm$2\\
375&&3.59&12&-\\
400&&3.58&10&7$\pm$2\\
\hline
\end{tabular}\\

\end{table}

Our XRD measurements show that the \tcn~becomes N deficient as
\Ts~is raised. This can also be seen among various works on
\tcn~available in the literature. For \Ts~= 440\,K, Oda $et$.
$al$. found LP = 3.586\,{\AA},~\cite{JMS87_Oda} Silva $et$. $al$.
and Wang $et$. $al$. found the LP = 3.54 and 3.56\,{\AA}~for \Ts~=
525\,K,~\cite{JAC15_Silva, Wang:CoN:TSF:09} at \Ts~= 625 and
725\,K, K. Ito. $et$. $al$. found LP =
3.52\,{\AA}.~\cite{2011_Co4N_K_Ito, JAP14_Ito} From here it
appears that a stoichiometric \tcn~phase has not been formed
before as in all works higher \Ts~was used. As we find that at
higher \Ts~even Co appears in the fcc phase and it's LP at
3.54\,{\AA}~is only about 5\p~lower than that \tcn, therefore fcc
Co has been mistaken for \tcn~phase. To further confirm this, we
measured N\pat~in our samples using SIMS.

\begin{figure} \center
\vspace{-1mm}
\includegraphics [width=70mm,height=80mm] {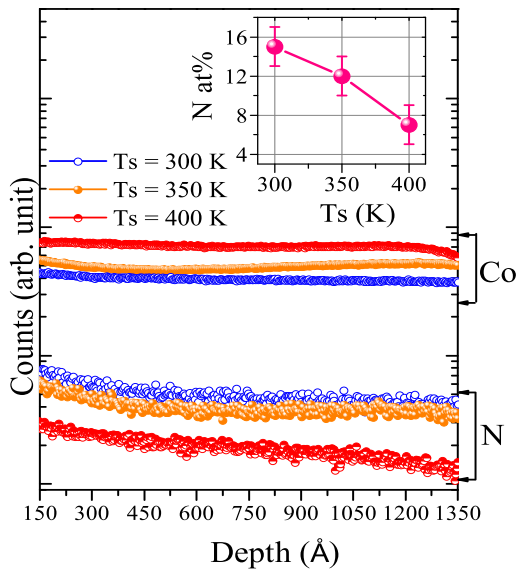}
\caption{\label{fig:sims} (Color online) SIMS depth profiles of
Co-N thin films deposited at \Ts~= 300, 350 and 400\,K. Inset
shows the variation of N\pat~as a function of \Ts.} \vspace{-1mm}
\end{figure}

SIMS depth profiles of Co and N for samples deposited at \Ts~=
300, 350 and 400\,K are shown in fig.~\ref{fig:sims}. We can see
that as \Ts~is increasing, N signal goes down while that of Co
goes up. This gives a direct evidence that N is diffusing out of
\tcn~even when \Ts~is just raised to 350\,K. We can see that our
Co profiles are almost constant throughout the depth of sample but
N profiles are somewhat skewed near surface regions. This may also
indicate that N is diffusing out from the surface as one would
expect, but large depth resolution of SIMS ($\sim$50-100{\AA}) and
poor signal does not allow to quantify N out diffusion. However,
by sandwiching a thin Co$^{15}$N marker layer between
Co$^\mathrm{nat}$N layers, we measured N self-diffusion and
results are presented in section~\ref{n_diff}.

Still N depth profiles can be used to estimate the N
concentration. Using a reference samples and following a procedure
described in earlier works,~\cite{JAC16_NPandey, JMMM17_NPandey,
AIP15_RG} we obtained N\pat, as shown in the inset of
fig.~\ref{fig:sims}. Since N profiles are somewhat skewed, an
average value was taken in the middle of the layer in all samples.
We can clearly see that N\pat~is decreasing almost linearly with
\Ts, in agreement with XRD results presented in table.~\ref{tab}.

\begin{figure} \center
\vspace{-1mm}
\includegraphics [width=70mm,height=35mm] {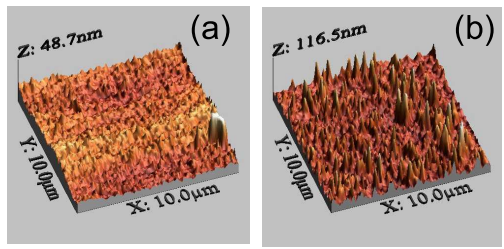}
\caption{\label{fig:afm} (Color online) 3D view of AFM images of
Co-N thin films deposited at \Ts~= 300\,K (a) and 400\,K (b).}
\vspace{-1mm}
\end{figure}

As N out diffusion takes place it may also affect the
microstructure. We can see from our AFM images
(fig.~\ref{fig:afm}) that the surface morphology of sample
deposited at \Ts = 300\,K is smoother than the one deposited at
400\,K. It appears that N out diffusion leaves behind a spike like
microstructure leading to larger surface roughness.

\begin{figure} \center
\vspace{-1mm}
\includegraphics [width=70mm,height=65mm] {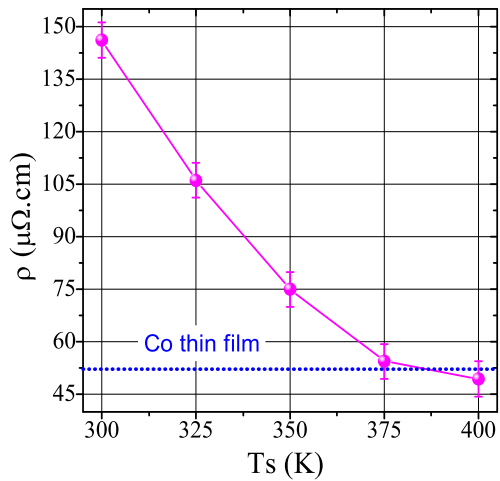}
\caption{\label{fig:res} (Color online) Dependence of electrical
resistivity ($\rho$) of Co-N thin films on \Ts.} \vspace{-1mm}
\end{figure}

We also measured the electrical resistivity ($\rho$) in our
samples and as can be seen in fig.~\ref{fig:res}, it decreases as
\Ts~increases. In earlier studies performed in the Co-N system, it
was demonstrated that $\rho$ maximizes with
N\pat.~\cite{JVSTA93_Shih, Vac01_Asahara, JVSTA04_Fang} Since in
our samples N\pat~is maximum for \Ts~= 300\,K sample, the value of
$\rho$ is also maximum and it decreases almost linearly as N\pat
decreases by increasing \Ts. It may be noted that for \Ts~= 400\,K
sample the value of $\rho$ becomes comparable to a pure Co thin
film. In agreement with other results, our electrical resistivity
also confirm that N is diffusing out when \Ts~is raised during
deposition of \tcn~thin films.

\begin{figure} \center
\vspace{-1mm}
\includegraphics [width=85mm,height=85mm] {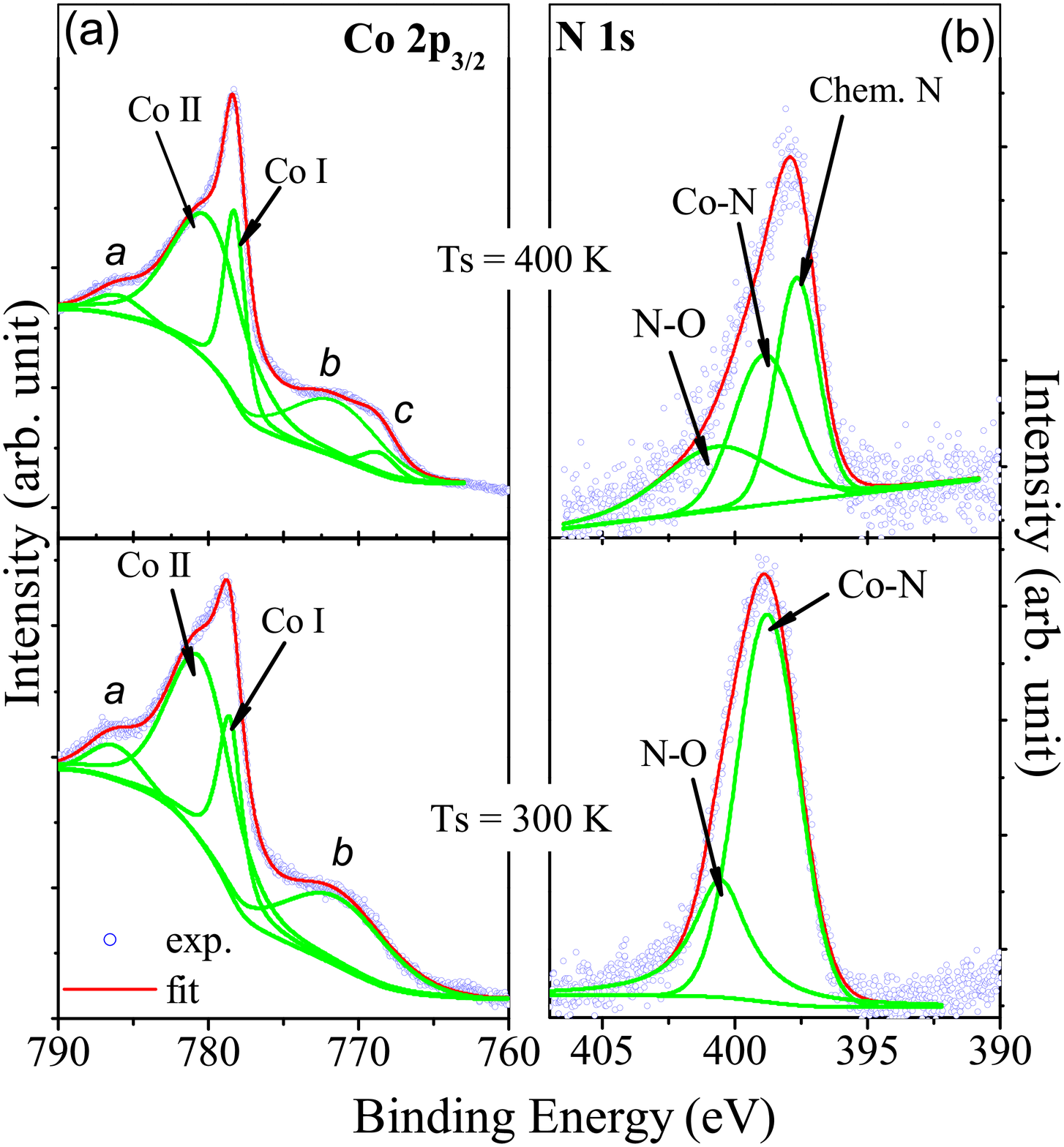}
\caption{\label{fig:xps} (Color online) Co 2p$_{3/2}$ (a) and N 1s
(b) XPS spectra of Co-N thin films deposited at \Ts~= 300 and
400\,K.} \vspace{-1mm}
\end{figure}

\subsection{\textbf{Electronic Structure}}
\label{elec}

We did XPS and XAS measurements to study the electronic structure
of our samples. Fig.~\ref{fig:xps}(a) and (b) shows  N 1s and Co
2p$_{3/2}$ spectra for samples deposited at \Ts~= 300 and 400\,K
and they were fitted using XPS peak fitting software. N 1s and Co
2p$_{3/2}$ spectra are significantly different for both samples.
In N 1s spectra, features corresponding to N-O (400.5\,eV) and
Co-N (398.6\,eV) can be seen in both samples. Presence of N-O
feature can be understood due to some surface oxidation (through
measurements were done after surface cleaning by sputtering) and
the Co-N feature stems from the bonding between Co and N. We find
that the area of the Co-N feature is about 4 times larger in \Ts~=
300\,K sample. On the other hand, a prominent feature for
chemisorbed N$_2$ (chem. N = 397.5\,eV) can only be seen in \Ts~=
400\,K sample. Weak Co-N and the presence of chem. N features
clearly signifies that some nitrogen has diffused out and
accumulated near surface regions (in agreement with SIMS data
shown in fig.~\ref{fig:sims}). Such signature of chem. N has also
been seen in \tfn~thin films after annealing in presence of N$_2$
gas.~\cite{SurfSci89_Diekmann} and also in other studies when N
diffuses out from metal nitrides.~\cite{SurfSci89_Diekmann,
SurfSci77_Kishi_absorbedN, Talanta78_Honda_chemisorbedN}

In order to fit the Co 2p$_{3/2}$ spectra, structural aspect of
\tcn~should also be considered. As shown in fig.~\ref{fig:fcc}, Co
sites (I and II) are different where Co II only has direct bonding
with N. We find that our Co 2p$_{3/2}$ can be fitted mainly with
these Co I and Co II components occurring at 778.5 and 780.5 \,eV,
respectively. We find that the fraction of Co I (Co II) increases
(decreases) when \Ts~is raised to 400\,K. This is in agreement
with N 1s spectra and can happen when the fraction of Co atoms
un-bonded with N increases. In addition, some features assigned as
`a', `b' and `c' can also be seen. Here `a' may be due to shake-up
transition~\cite{Co_shakeup_XPS_Frost} which is usually observed
in the metals while features `b' and `c' are due to the L$_3$VV
Auger transition.~\cite{JESRP12_Powell_Auger}

\begin{figure} \center
\vspace{-1mm}
\includegraphics [width=70mm,height=90mm] {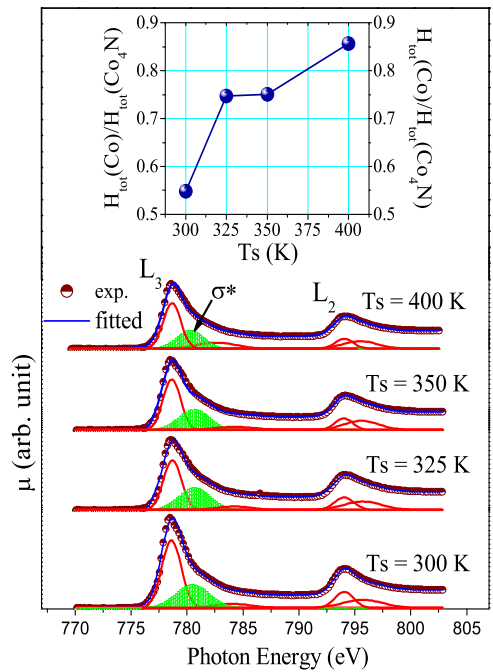}
\caption{\label{fig:xas} (Color online) Co L edge XANES spectrum
of Co-N samples deposited at \Ts~= 300, 325, 350 and 400\,K. Inset
shows the variation in H$_\mathrm{tot}$(Co)/H$_\mathrm{tot}$(\tcn)
with \Ts, here error bars are typically the size of symbols.}
\vspace{-1mm}
\end{figure}

Fig.~\ref{fig:xas} show the Co L-edge XAS spectra of \tcn~thin
films deposited at \Ts~= 300, 325, 350 and 400\,K. Two main
spectral features were observed around 778\,eV~(L$_{3}$) and
794\,eV (L$_{2}$). They arise due to well-known spin-orbit
splitting of the Co 2p core-hole
(2p$^6$3d$^n$$\rightarrow$2p$^5$3d$^{n+1}$) transition. We used
Athena software~\cite{Athena} to fit XAS spectra using a step and
Gaussian function after pre and post-edge normalization. Co
L$_{3}$ edge can be deconvoluted into three ($\sim$ 778, 780,
784\,eV) and L$_2$ into two (794 and 795\,eV) sub spectra. Here
the peaks at $\sim$778 and 784\,eV correspond to the Co I site in
\tcn. On the other hand, the peak at $\sim$780\,eV~(shown as
shaded region in fig.~\ref{fig:xas}) corresponds to $\sigma$$^*$
anti-bonding which occurs due to a strong hybridization of `pd'
states between Co II and N atoms in \tcn.~\cite{APL_11_K_Ito_Co4N,
JAP15_KIto_Fe4N} We can clearly see that the $\sigma$$^*$ feature
becomes weak as \Ts~is increasing indicating reduced hybridization
between Co and N. From the fitting of this peak, hybridization
between Co and N can be quantified. Compared to \Ts~= 300\,K
sample, it reduces by about 13, 21 and 38\p~as \Ts~is raised to
325, 350 and 400\,K, respectively.

In agreement with other results, XAS measurements also confirm
that the fraction of Co atoms unbonded with N increases as \Ts~is
increased. This can be further confirmed by directly comparing the
overall intensities of L-edges. It is known that the sum of
amplitudes of L$_{3}$ and L$_{2}$ edges in XAS is proportional to
number of holes (H$_\mathrm{tot}$ = HL$_{3}$ + HL$_{2}$, where
HL$_{3}$, HL$_{2}$ is the amplitude of L$_3$ and L$_2$ edges)
present in $d$ state of the metal.\cite{PRB85_Morrison_XAS,
PRB87_Morrison_XAS} Comparing H$_\mathrm{tot}$(\tcn) for Co-N
samples together with H$_\mathrm{tot}$(Co) for a pure Co film can
yield variances between them as for N$\rightarrow$0,
$\frac{H_\mathrm{tot}(Co)}{H_\mathrm{tot}(Co_4N)}$$\rightarrow$1.
As shown in the inset of fig.~\ref{fig:xas},
H$_\mathrm{tot}$(Co)/H$_\mathrm{tot}$(\tcn) increases from 0.55 to
about 0.85 when the \Ts~is raised from 300 to 400\,K. Since later
is approaching to unity confirming that the amount of N is
becoming negligible due to N out-diffusion.

Both XPS and XAS measurements unambiguously confirm the phase
formation of \tcn~at \Ts~= 300\,K and when it is raised to 400\,K,
substantial amount of N diffuses out leaving behind mainly fcc Co.

\subsection{\textbf{Magnetic properties}} \label{mag}

\begin{figure} \center \vspace{-1mm}
\includegraphics [width=85mm,height=85mm] {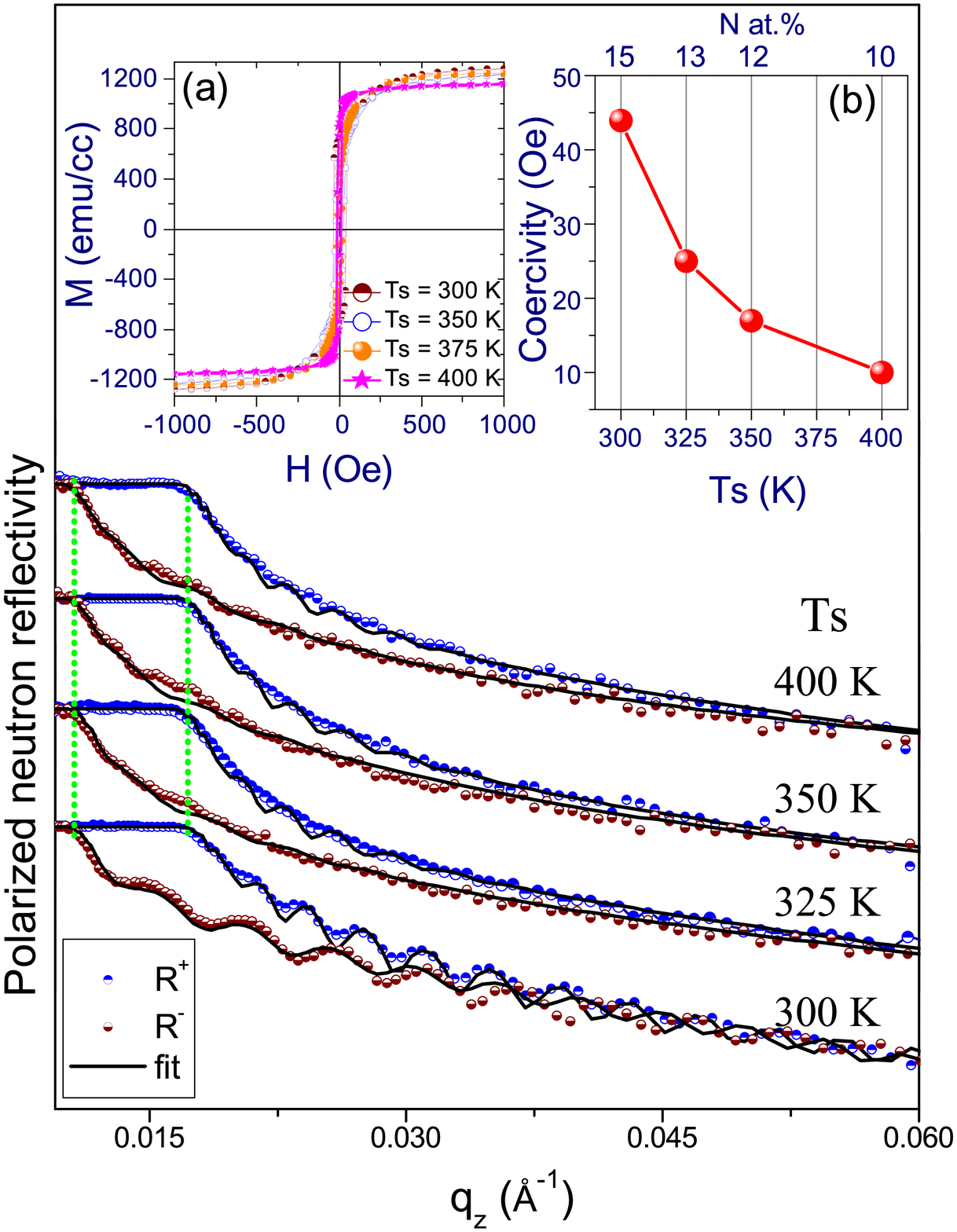}
\caption{\label{fig:pnr} (Color online) Fitted PNR spectra of Co-N
thin films deposited at \Ts~= 300, 325, 350 and 400,K. Inset
showing the M-H hysteresis loop (a) and variation of coercivity of
these samples as a function of \Ts~(b). Typical errors in the
value of coercivity is the size of symbol used here.}
\vspace{-1mm}
\end{figure}

To measure the saturation magnetization ($M$) in our samples, bulk
magnetization (SQUID-VSM) and PNR measurements were carried out.
It may be noted that mostly bulk magnetization measurements have
been used to deduce $M$ in \tcn~thin films. Due to ambiguities in
estimating sample volume and density, there may be errors in
estimating correct $M$, which is evident from large variations
observed (between 46 to 150 emu/gm and 485 - 1300 emu/cc) in
\tcn~thin films.~\cite{JMS87_Oda, JAC15_Silva, 2011_Co4N_K_Ito,
TSF14_Silva, Wang:CoN:TSF:09, MSEB08_Jia, ASS17_Li_Zhang_Co4N} On
the other hand since in PNR, the sample volume and substrate
demagnetization effects do not play a role in measurement of $M$,
we did PNR for precise measurement of $M$ in our samples.
Fig.~\ref{fig:pnr} shows PNR and M-H loops of samples deposited at
different \Ts. The inset (a) of fig.~\ref{fig:pnr} shows M-H loops
and the inset (b) compares the coercivity (Hc). We find that Hc is
small indicating that samples are soft magnetic and it almost
decreases linearly as \Ts~increases.

From the M-H loops, we find that $M$ = 1304, 1286, 1262 and
1174($\pm$20)\,emu/cc for \Ts~= 300, 325, 350 and 400\,K,
respectively. PNR patterns of our samples are shown in
fig.~\ref{fig:pnr}. From the fitting of these (using
GenX~\cite{GenX}), we find that $M$ =
1.73$\pm0.05$\,$\mu_\mathrm{B}$/Co atom for \Ts~= 300\,K sample
while for \Ts~= 325, 350 and 400\,K samples it comes out to be
about 1.65$\pm$0.05\,$\mu_\mathrm{B}$/Co atom. Clearly both S-VSM
and PNR measurements suggest that the value of $M$ is the largest
for \Ts~= 300\,K sample and decreasing as \Ts~is increasing. This
behavior can be understood from the well-known high-volume
high-moment derived for tetra metal nitrides.~\cite{PRB07_Matar,
JMMM10_Imai} An empirical relation for a binary alloy (consisting
of elements A and B) between LP and $M$ is given
by:~\cite{AIPConf74_Shiga_MVE}

\begin{equation}
\label{equ:lp_m} LP = a_A(1-x) + a_Bx + CM
\end{equation}

Where, $a_A$, $a_B$ and $C$ are the constant like parameters and
$x$ is the atomic fraction. However, the exact quantification of
$M$ is not possible using eq.~\ref{equ:lp_m} due to ambiguities in
exact calculation of constants.~\cite{AIPConf74_Shiga_MVE}
Alternatively, using a rather simplistic approach based on the
atomic theory model and considering the dependence of LP on the
composition of \tcn~as shown in fig.~\ref{fig:lp_n}, we can
calculate $M$ for samples deposited at different \Ts.

\begin{table}  \vspace{-5mm}
\caption{\label{table2} Magnetic configuration of Co I and Co II
sites in \tcn.}
\begin{tabular}{cccccc} \hline
Co site&Coordinate&d occupancy&$\mu$&$nu$&$nd$\\
\hline
I&(0,0,0)&3d$^8$&2&0&1\\
II&(0,1/2,1/2)&3d$^7$&3&1&0\\
II&(1/2,0,1/2)&3d$^7$&3&1&0\\
II&(1/2,1/2,0)&3d$^7$&3&1&0\\
\hline
\end{tabular}
\end{table}

By considering the postulates of atomic theory that (i) filling of
electrons according to Hund's rule (ii) overlapping of nearest
neighbor will cause antiparallel alignment of 3d shell. In the
\tcn~structure Co I and Co II sites (as shown in
fig.~\ref{fig:fcc}) contribute in determination of $M$ as:
~\cite{JOM55_Wiener, PhyR58_Frazer}

\begin{equation}
\label{equ:atm} M = (nu_{\mathrm {II}}-nd_{\mathrm
{II}})\mu_\mathrm{II} + (nu_{\mathrm {I}}-nd_{\mathrm
{I}})\mu_\mathrm{I}
\end{equation}

Here, $nu_{\mathrm {II}}$ and $nd_{\mathrm {II}}$ are number of Co
II sites having spin up and down electrons, respectively.
Similarly $nu_{\mathrm {I}}$ and $nd_{\mathrm {I}}$ can be
interpreted for Co I. $\mu_\mathrm{I}$ and $\mu_\mathrm{II}$ is
the resultant moment of Co I and Co II, respectively. Magnetic
configuration of these sites is given in table~\ref{table2}. Using
eq.~\ref{equ:atm} and table~\ref{table2}, we get $M$ =
(3-0)3+(0-1)2 = 7\,$\mu_\mathrm{B}$ per f.u., yielding $M$ =
1.75\,$\mu_\mathrm{B}$/Co atom for stoichiometric \tcn. This is in
agreement with theoretical calculations based on DFT
calculations.~\cite{PRB07_Matar, JAC14_Imai} Considering a similar
approach described in section~\ref{smc}, in a supercell having 8
unit cells of \tcn~and by sequentially taking out one N atom, we
get $M$ = 1.74, 1.73 and 1.72\,$\mu_\mathrm{B}$/Co atom for LP =
3.65, 3.63 and 3.57\,{\AA} corresponding to N\pat~= 15, 13 and
10\p, respectively. Broadly these theoretical calculations are in
agreement with our experimental results.

\subsection{\textbf{Target poisoning affecting \tcn~phase formation}}\label{tp}
From the above results, we have shown that \tcn~thin film attains
the highest value of LP as \Ts~is lowered and N\pat~also
increases. We find that LP was maximum for \Ts~= 300\,K sample at
3.65(5)\,{\AA} and the amount of N in \tcn~was 15$\pm2$\p. Both
these are somewhat smaller than the theoretical values: for
stoichiometric \tcn, LP should be 3.735\,{\AA} and N concentration
at 20\pat.~\cite{PRB07_Matar} Since further lowering of \Ts~may
produce a disordered phase, other parameters that could influence
stoichiometry of \tcn~film were considered.

\begin{figure} \center
\vspace{-1mm}
\includegraphics [width=80mm,height=80mm] {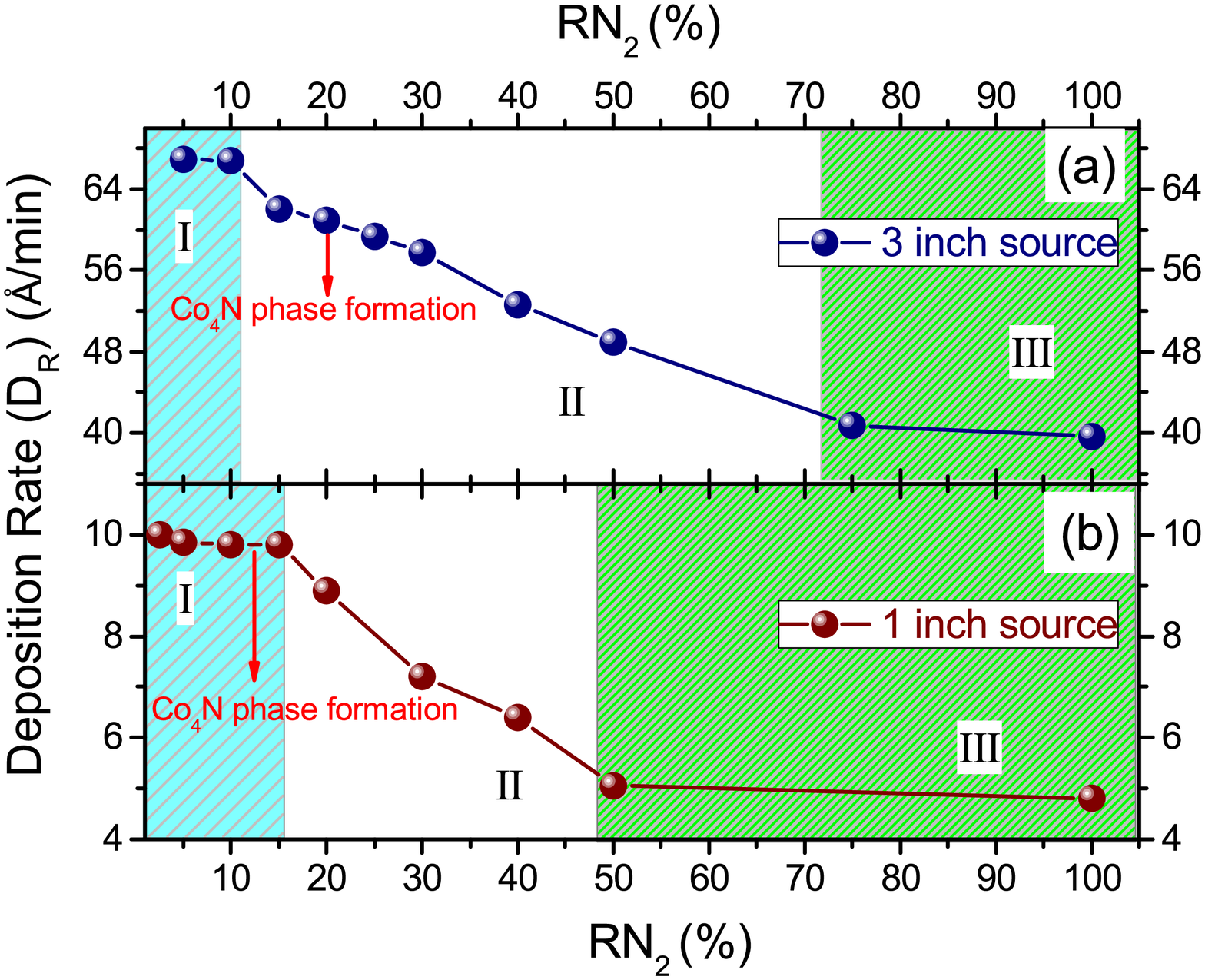}
\caption{\label{fig:rate} (Color online) Variation of deposition
rate with \pn~in Co-N thin films deposited using 3 (a) and 1\,inch
(b) targets. Here region `I' corresponds to the `metallic state'
of target, region `II' denotes to the `transition' and `III' to
the `poisoned state' of Co target. Errors in deposition rates are
lower than the size of symbols used here.} \vspace{-1mm}
\end{figure}

In a reactive sputtering process, variations in the deposition
rate (D$_R$) with the amount of reactive gas is a key for phase
formations. When the D$_R$ is similar to elemental metal target,
it is referred as `metallic state' and when D$_R$ become very
small (compared to metallic state), target is in `poisoned state'
due to compound formation taking place at the target itself. In
between these two states is a `transient state' which is rather
ill-defined. When a compound is formed from such `transient
state', it is affected by so called hysteresis
effects.~\cite{TSF14_Upgraded_berg, Berg_model_TSF05,
APL05_Nyberg_RS, AIP07_Andres_HiPIMS, RS10_Review, RS08_Kubart,
RS06_Kubart} We measured D$_R$ by depositing a series of Co-N thin
films at \Ts~= 300\,K by systematically increasing \pn~during
deposition. Thicknesses were measured using x-ray reflectivity and
obtained D$_R$ are shown in fig.~\ref{fig:rate} (a). We find that
for \pn~= 0-10\p, the target is in `metallic state' (region I),
between 15-50\p~it is in `transient state' (region II) and above
this \pn~values, it attains the `poisoned state' (region III).
Since the \tcn~phase was formed for \pn~= 20\p, it emerges from
the `transient state' of the target and therefore even a slight
variation in \pn~could vary the composition.

A lot of efforts have been made to produce hysteresis free films
in reactive sputtering process. Some of efforts realize on
enhancing the ionic concentration by doing high power impulse
magnetron sputtering.~\cite{AIP07_Andres_HiPIMS} A novel approach
within the dc magnetron sputtering process was shown for aluminium
oxides by reducing the target erosion area. This essentially
reduces the consumption of reactive gas in target poisoning and
extending the `metallic state' of the
target.~\cite{APL05_Nyberg_RS} These effects were also
demonstrated in theoretical models proposed by Berg et
al.~\cite{TSF14_Upgraded_berg, Berg_model_TSF05}

\begin{figure} \center
\vspace{-1mm}
\includegraphics [width=80mm,height=65mm] {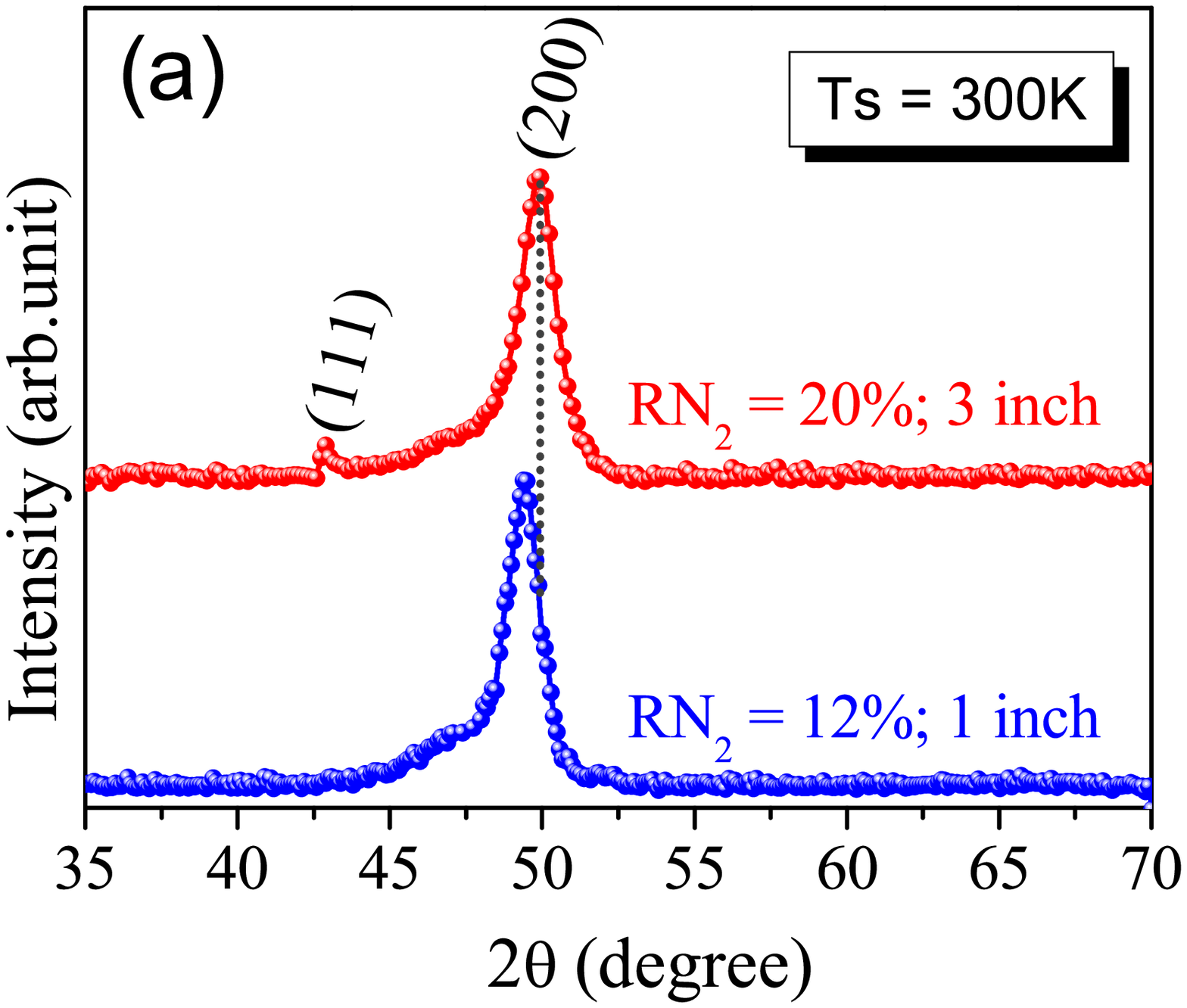}
\includegraphics [width=85mm,height=60mm] {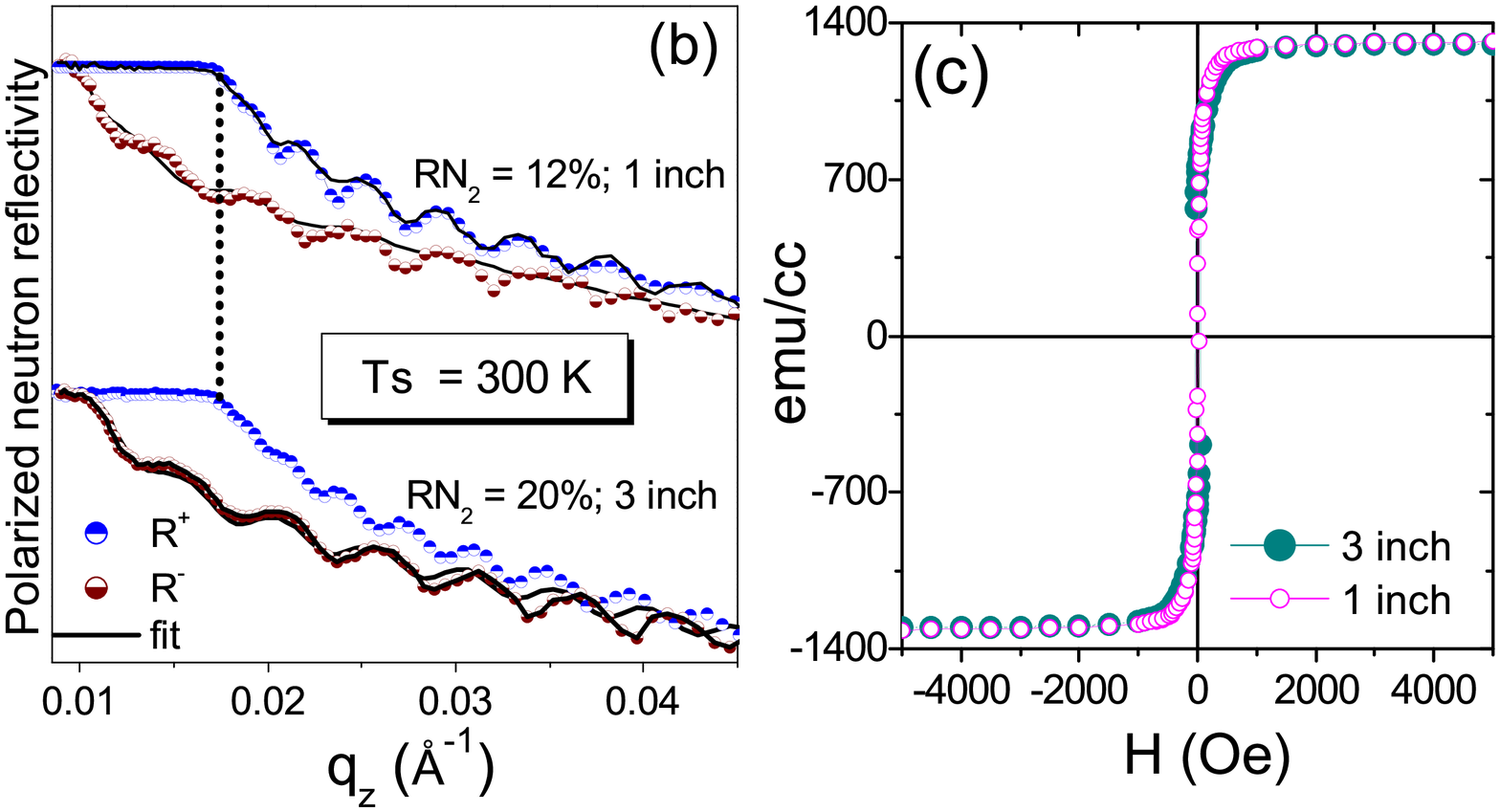}
\caption{\label{fig:xrd_1inch} (Color online) XRD patterns Co-N
thin films deposited at \pn~= 12 and 20\p~using 1 and 3\,inch Co
targets, respectively both at \Ts~= 300\,K (a), fitted PNR
patterns (b) and M-H hysteresis curves of these thin films (c).}
\vspace{-1mm}
\end{figure}

Since in our case when we use a 3\,inch diameter Co target, the
\tcn~phase was forming from the `transient state' and it was under
stoichiometric, we attempted to prepare the \tcn~phase from a
smaller 1\,inch diameter Co target at \Ts~= 300\,K. As can be seen
from fig.~\ref{fig:rate} (b), the `metallic state' has now been
extended and the `transient state' has narrowed down but the
`poisoned state' has also been extended. This implies that the
smaller target remains rather insensitive to the reactive gas flow
in the beginning but it becomes even more sensitive (than larger
target) to the reactive gas at higher flows of reactive gas. This
may happen as ion implantation of reactive gas becomes more
dominant in the smaller area source. This behavior is in agreement
with modified Berg's model.~\cite{TSF14_Upgraded_berg}

From the 1\,inch Co target, we find that \tcn~phase is forming
already at \pn~= 12\p~(within the `metallic state') as shown in
fig.~\ref{fig:xrd_1inch} (b). A comparison of this sample together
with \tcn~sample deposited with 3\,inch source clearly shows that
the (200) XRD peak is shifted to lower angle and the obtained
value of LP is 3.69(5)\,{\AA} which is very close to its
theoretical value and is the highest value obtained so far for
\tcn~thin films. Our SIMS measurements (not shown here) also
confirmed that N concentration is about 18$\pm$2\p. Clearly by
reducing the target area, optimizing \Ts~we could deposit the
stoichiometric \tcn~phase matching well with its theoretical
values. However, it may be noted that overall deposition rates
decrease significantly when the smaller source was used.

Magnetization of this sample was also measured and PNR and MH
curves are compared for \tcn~thin films deposited using 1 and
3\,inch sources as shown in fig.~\ref{fig:xrd_1inch} (b, c). From
PNR we can see that the R$^+$ is slightly shifted to higher q$_z$
values in \tcn~thin film deposited using 1\,inch source indicating
that $M$ is higher here, although it is not as noticeable from the
M-H loop. From the fitting of PNR data we obtained $M$ =
1.75$\mu_\mathrm{B}$/Co atom which is slightly higher than the
\tcn~sample deposited with the 3\,inch Co source. These results
are can be understood due to a small rise in LP of \tcn~film when
deposited with the 1\,inch Co source.

\subsection{\textbf{Epitaxial \tcn~film}}\label{epi}

\begin{figure} \center
\vspace{-1mm}
\includegraphics [width=85mm,height=60mm] {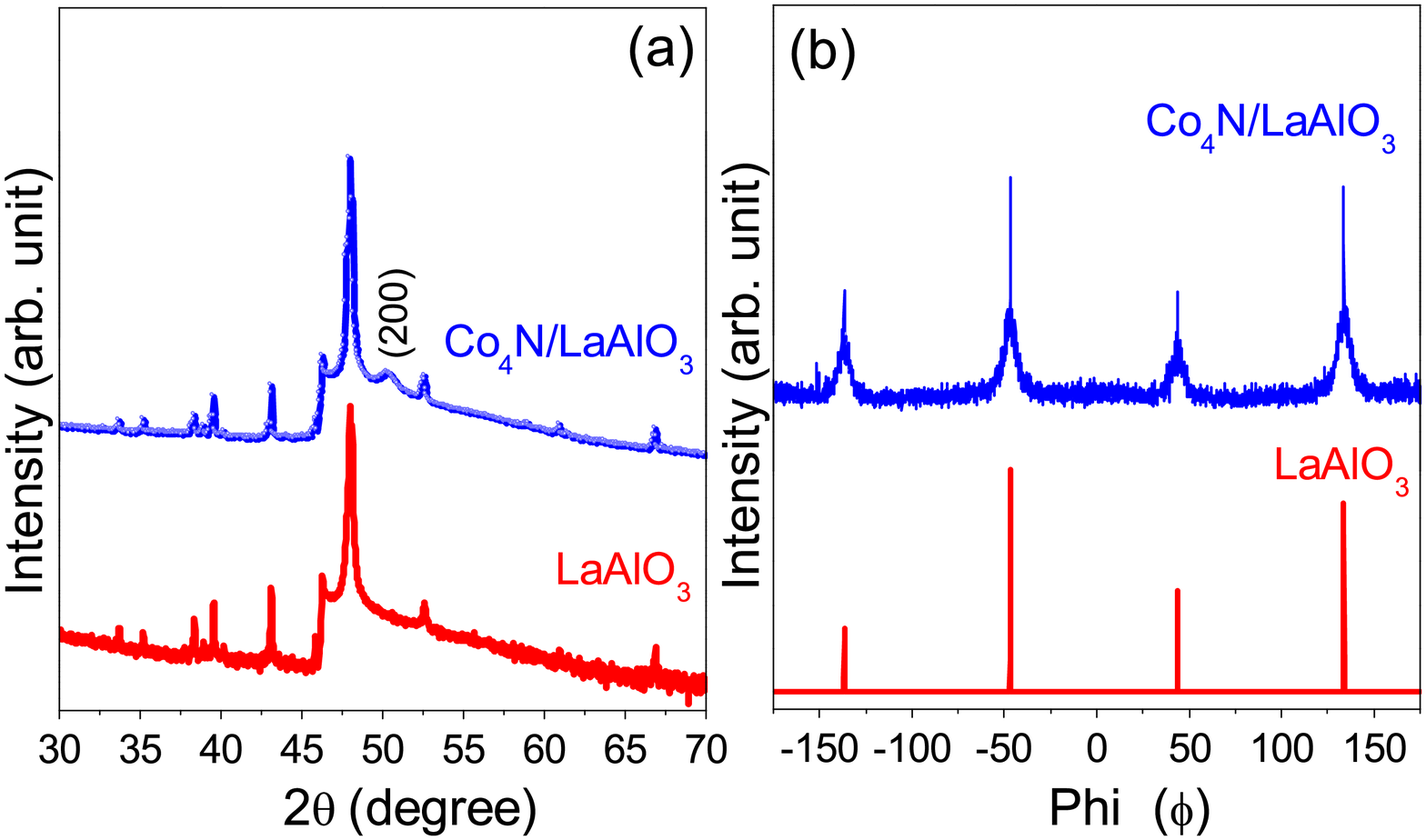}
\includegraphics [width=90mm,height=53mm] {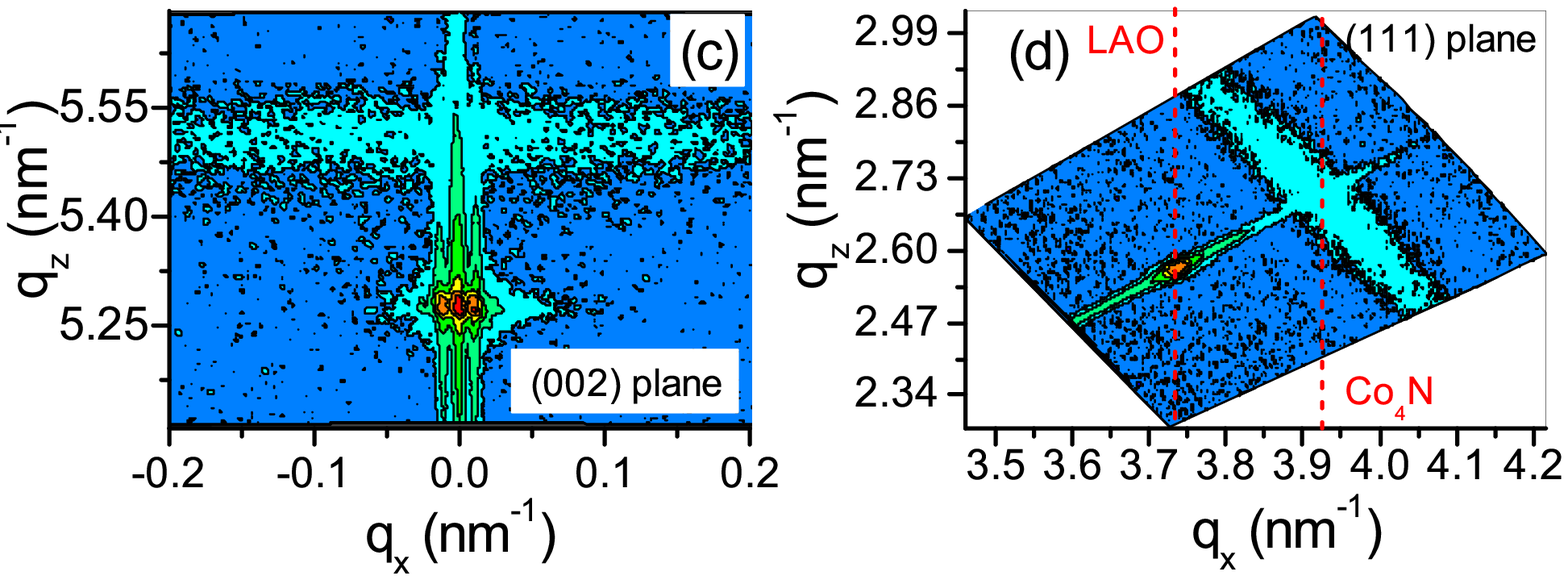}
\caption{\label{fig:epi_xrd} (Color online) XRD spectra (a) and
Phi scan along (100) plane (b) of \tcn~thin film grown on LAO
(100) substrate. Reciprocal space mapping of \tcn~thin film
deposited at LAO (100) substrate (c).} \vspace{-1mm}
\end{figure}

So far, we have discussed about finding pathways for a
stoichiometric \tcn~film and got its LP up to about 99\p~of the
theoretical value. This was achieved by depositing samples at
\Ts~= 300\,K and using the `metallic state' in a 1\,inch Co target
during the reactive sputtering process. These optimized conditions
were applied to deposit a \tcn~thin film on a LaAlO$_3$ (LAO)
substrate oriented along (100) plane. Since LAO has a very small
lattice mismatching with \tcn~($\sim$1.3\p~), it is most suitable
to grow an epitaxial \tcn~thin film. As a reference, a
polycrystalline sample was again deposited on a quartz (SiO$_2$)
substrate together with that on LAO. Fig.~\ref{fig:epi_xrd} (a)
shows XRD pattern of \tcn~film deposited on LAO and quartz
substrate and for comparison XRD pattern taken on the bare LAO
substrate is also shown. In agreement with
fig.~\ref{fig:xrd_1inch}(a), the \tcn~sample on the quartz
substrate is identical but on the LAO substrate has (200) plane at
a slightly higher angle. The LP for the \tcn~film on LAO substrate
comes out to be 3.635\,{\AA}.

To confirm the epitaxial nature of the sample deposited on LAO
substrate, Phi scan along (100) plane (fig.~\ref{fig:epi_xrd} (b))
and RSM measurements were carried out. Here the Phi scan of
substrate shows a very clear four fold reflections which occur in
cubic symmetry. Similarly, the reflections from film are also
completely overlapping the substrate peaks, conforming the
epitaxial growth of \tcn~thin film on LAO substrate.

Fig.~\ref{fig:epi_xrd} (c) and (d) shows RSM scans for this sample
along (002) and (111) planes, respectively. From here we deduced
out-of-plane (002) and in-plane (111) LP that comes out to be
3.635\,{\AA} and 3.71\,{\AA}, respectively. Since the film on the
quartz substrate was found to be fully stoichiometric \tcn~having
LP$\sim$3.7\,{\AA}, the film on the LAO substrate is also expected
to be stoichiometric. The discrepancy in LP for \tcn~film
deposited on LAO substrate may arise if it has a tensile strain
along in-plane and a compressive strain along the out-of-plane
direction. If this strain stems from a lattice mismatch between
film and substrate, the broken lines shown in
fig.~\ref{fig:epi_xrd} (d) should have appeared at similar q$_x$
values.~\cite{JAP02_RSM_strain_relax, JCG12_RSM_strain_relaxed}
But since they are well-separated, the strain could be within the
film itself leading to different values of in-plane and
out-of-plane LP. In addition, it may be noted that the spot
corresponding to \tcn~film in fig.~\ref{fig:epi_xrd} (d) is quite
broad and also the width of Phi reflections in
fig.~\ref{fig:epi_xrd} (b) is large. Both these indicate a mosaic
type epitaxial growth.~\cite{JAP02_RSM_strain_relax,
JCG12_RSM_strain_relaxed} This could be due to the fact that at
low \Ts~(300\,K) adatoms would not have sufficient energy to
establish the long range ordering to its fullest.

\begin{figure} \center
\vspace{-1mm}
\includegraphics [width=70mm,height=75mm] {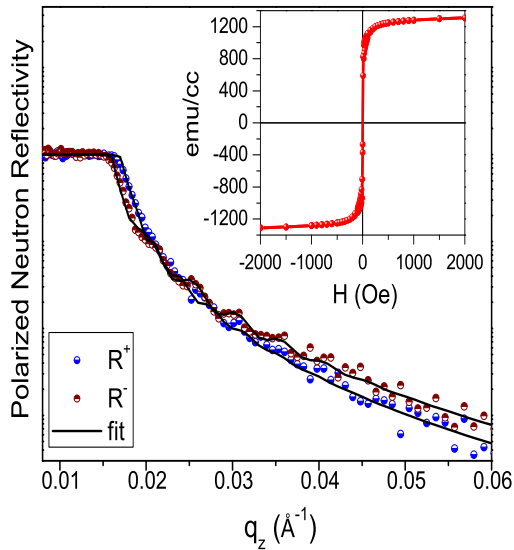}
\caption{\label{fig:epi_m} (Color online) Fitted PNR spectra of
epitaxial \tcn~thin film on LAO substrate. Inset showing the M-H
hysteresis loop of the same sample.} \vspace{-1mm}
\end{figure}

The PNR and M-H loop for the epitaxially grown \tcn~thin film are
shown in fig.~\ref{fig:epi_m}. Since the scattering length density
of LAO substrate is large, the splitting between R$^+$ and R$^-$
is not as appreciable as found for the polycrystalline sample
deposited on quartz (SiO$_2$) substrate. From the M-H loop $M$
comes out to be $\sim$1320\,emu/cc and from PNR measurements, it
is $\sim$1.75$\mu_\mathrm{B}$/Co atom. Both these values are
matching well with polycrystalline \tcn~sample deposited under
similar conditions.

\begin{figure} \center
\vspace{-1mm}
\includegraphics [width=60mm,height=50mm] {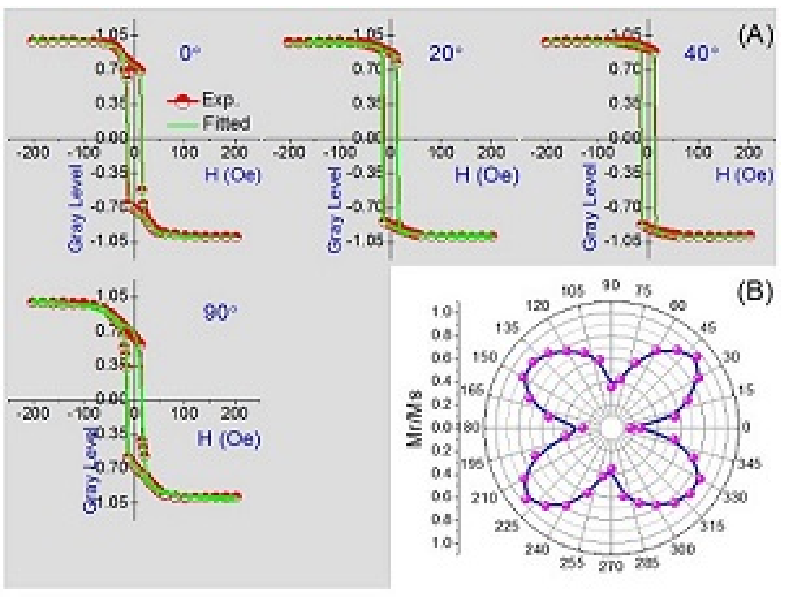}
\includegraphics [width=85mm,height=35mm] {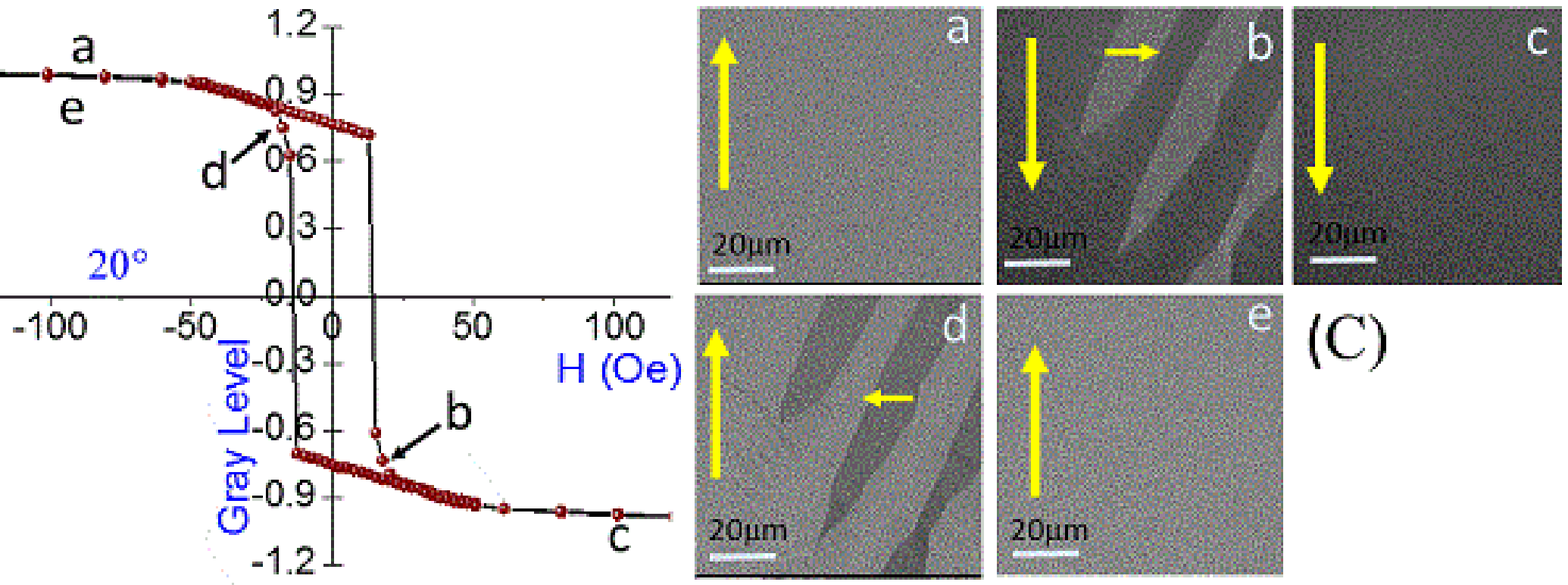}
\caption{\label{fig:moke} (Color online) Longitudinal MOKE
hysteresis loops of epitaxial \tcn~thin film (A). Polar plot of
squareness (Mr/Ms) and coercivity with applied field angle (B).
Kerr domain evolution images of epitaxial \tcn~thin film for
20$^{\circ}$ hysteresis loop (C).} \vspace{-1mm}
\end{figure}

The influence of epitaxy on the magnetic property was investigated
from MOKE and Kerr microscopy measurements e.g. the presence of
cubic symmetry should appear in the polar plot of squareness (S =
Mr/Ms; here Mr and Ms are remanence and saturation magnetization)
with the applied field angles. This has been amply demonstrated
for \tfn~thin films but not yet for \tcn.~\cite{PRB04_Fe4N_Costa}
We measured hysteresis loops (in longitudinal MOKE setup) by
changing the angle of applied magnetic field at an interval
10$^{\circ}$ between 0 and 180$^{\circ}$ assuming similar for
other two quadrants. They are shown in fig.~\ref{fig:moke} (A) for
representative angles of 0, 20, 40 and 90$^{\circ}$. Further, we
can clearly see presence of two loops and to quantify them, they
were fitted using:~\cite{JAP94_twoMH_loop_Stearns,
PRB14_AT_FedopN}

\begin{equation}
\label{equ:mh_fit} M(H) = \sum\limits_{i}\frac{2Ms^i}{\pi}
[(\frac{H{\pm}Hc^i}{Hc^i}) tan(\frac{\pi S^i}{2})]
\end{equation}

From the fitting, we obtain two magnetic contributions having
coercivity, Hc $\sim$ 15 and 40\,Oe and their corresponding amount
comes out to be about 94 and 6\p, respectively. Here, it appears
that majority fraction should be due to \tcn~phase and the minor
one could be due to some Co impurity. In addition, the fitting of
these loops yields variation of the $S$ as a function of applied
field angles shown in the polar plot fig.~\ref{fig:moke} (B). The
polar plot clearly shows an in-plane biaxial magnetic anisotropy
(four fold). As expected, we can see that easy and hard
magnetization axes are along (100) and (110) direction,
respectively. Such four fold anisotropy is expected due to cubic
symmetry and confirm the epitaxial growth of \tcn~as evidenced
earlier for epitaxial \tfn~thin films.~\cite{PRB04_Fe4N_Costa} In
agreement with XRD measurements, our MOKE measurements also
confirm the epitaxial nature of \tcn~thin films deposited on LAO
substrate. Here the magnetocrystalline constant (K) can be
estimated assuming that the anisotropy field is nearly equal to
the saturation field and comes out to be
$\sim$1.2$\times$10$^4$\,J/m$^3$ which is smaller than that of Co
($\sim$10$^5$\,J/m$^3$) but close to \tfn.~\cite{PRB04_Fe4N_Costa}
Furthermore, a reversal process by 90$^{\circ}$ domain wall
nucleation should take place when the magnetic field is applied in
between easy and hard axes.~\cite{PRB04_Fe4N_Costa} This can be
evidenced in Kerr microscopy images as shown in
fig.~\ref{fig:moke}(C). Images were captured at points a(=e), b,
c, and d and shown there for an applied field angle of
20$^{\circ}$. The 180$^{\circ}$ magnetization reversal can be
clearly seen from the image $a$ to $e$ followed by two consecutive
90$^{\circ}$ domain wall nucleation in image $b$ and $d$ (shown by
arrow).

\subsection{\textbf{N Self Diffusion in Co-N}}\label{n_diff}
\begin{figure} \center
\vspace{-1mm}
\includegraphics [width=70mm,height=85mm] {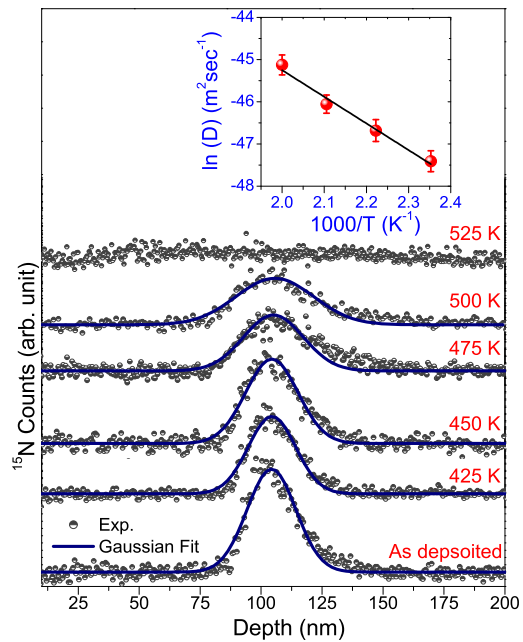}
\caption{\label{fig:15N_diff} (Color online) SIMS depth profile of
$^{15}$N on a Si
(substrate)[Co$^\mathrm{nat}$N(100\,nm)$|$Co$^{15}$N(2\,nm)$|$Co$^\mathrm{nat}$N(100\,nm)]
trilayer after annealing at different temperatures. Inset shows
the Arrhenius plot of diffusivity of N.} \vspace{-1mm}
\end{figure}

The results obtained from this work can be understood if N is
diffusing out from Co-N even if we raise the \Ts~above 300\,K.
Signatures of such N out-diffusion can be clearly seen in the N
depth profiles shown in fig.~\ref{fig:sims}. However, to quantify
it, N self-diffusion should be measured. We find that N
self-diffusion in the Co-N system has not been measured before.
Following a similar approach adopted for the FeN thin
films,~\cite{PRB15_AT_FeN, JAP11_MG_AT, PRB02_MG_FeN} we prepared
a sample
Co$^\mathrm{nat}$N(100\,nm)$|$Co$^{15}$N(2\,nm)$|$Co$^\mathrm{nat}$N(100\,nm)
on a Si substrate at \Ts~= 300\,K and under identical conditions
that were used to prepare the FeN sample.~\cite{PRB15_AT_FeN,
JAP11_MG_AT} For determining the diffusivity of N, SIMS depth
profiles were measured after isochronal thermal annealing at
different temperatures for 1\,hour at each temperature and are
shown in fig.~\ref{fig:15N_diff}. Due to thermal annealing,
$^{15}$N profiles get broadened and the width of this peak has
been used to deduce the diffusivity:~\cite{PRB15_AT_FeN,
JAP11_MG_AT}

\begin{equation}\label{diff}
D(t) = \frac{\sigma_t^2-\sigma_\circ^2}{2t}
\end{equation}

Where, D(t) is the time-averaged diffusion coefficient and
$\sigma_t$ is the standard deviation (before annealing; t = 0 and
after an annealing time t) obtained after fitting $^{15}$N
profiles to a Gaussian function as shown in
fig.~\ref{fig:15N_diff}. Using eq.~\ref{diff}, diffusivity ($D$)
of N has been measured at different temperatures and follows an
Arrhenius type behavior given by:

\begin{equation}\label{arh}
D = D_\circ~\mathrm{exp}(-E/k_{B}T).
\end{equation}

Here, $D_\circ$ denotes the pre-exponential factor, $E$ the
activation energy, T annealing temperature, and $k_B$ is the
Boltzmann's constant. Using eq.~\ref{arh} we get a straight line
as shown in inset of fig.~\ref{fig:15N_diff}, yielding activation
energy $E$ = 0.5$\pm$0.05\,eV. Here it may be noted that this
activation energy is valid only for the isochronal annealing
treatment and relaxation of diffusivity was not taken into
account. We can now directly compare the diffusivity of N in FeN
together with that in CoN. In FeN we find that $^{15}$N profile
can still be seen when the sample was annealed at
600\,K~\cite{PRB15_AT_FeN} whereas, it completely disappears above
500\,K as shown in fig.~\ref{fig:15N_diff}. Also the activation
energy for N self-diffusion in FeN is typically about
1.5\,eV.~\cite{PRB15_AT_FeN, JAP11_MG_AT} Both poorer thermal
stability and smaller activation energy is an indication that N
self-diffusion in CoN would be significantly higher than that in
FeN. Therefore, when CoN samples are deposited even at a moderate
temperature like 400\,K, N diffuses out leaving behind fcc Co that
has been often mistaken for \tcn. High self-diffusion of N in Co-N
can also be understood from the fact that the enthalpy of nitride
formation in the Co-N system is typically $\sim$
0\,J/mol,~\cite{JAC14_Imai} whereas it is around
-47\,kJ/mol~\cite{Tessier_SSS00} in the Fe-N system. This as well
explains the poorer thermal stability cobalt nitrides as compared
to iron nitrides.

\section{Conclusion}
\label{con}

In this work, we investigated the pathways to attain a
stoichiometric \tcn~phase. This could be achieved when the
\tcn~phase emerges from the metallic state of Co target. More
importantly, the \Ts~has an immense role in \tcn~phase formation.
Contrary to general assumption, the \tcn~phase can only be formed
at \Ts~= 300\,K. Detailed analysis of structure, composition and
magnetic properties of \tcn~samples deposited at different
\Ts~reveals that they become N deficient even if the \Ts~is raised
to 400\,K. This is an important observation as most of the
\tcn~films reported so far were deposited at \Ts$>$400\,K and the
LP of resulting films was more closer to fcc Co rather than \tcn.
We show that stoichiometric polycrystalline and epitaxial
\tcn~films can be formed at \Ts~= 300\,K. These results have been
explained by measuring N self-diffusion in CoN. We found that N
self-diffusion comes out to be orders of magnitude larger than the
anticipated value. This explains that while the \tfn~can be formed
at high \Ts~($\sim$670\,K), \tcn~can only be formed at low \Ts.

\section*{Acknowledgments}
Authors thank the Department of Science and Technology, India
(SR/NM/Z-07/2015) for the financial support and Jawaharlal Nehru
Centre for Advanced Scientific Research (JNCASR) for managing the
project. A part of this work was performed at AMOR, Swiss
Spallation Neutron Source, Paul Scherrer Institute, Villigen,
Switzerland. We acknowledge help received from L. Behera and Seema
in sample preparation and various measurements. M. Horisberger is
acknowledged for depositing $^{15}$N enriched CoN samples. We are
thankful to A. Gome for XRR, V. Ganesan and M. Gangrade for AFM,
R. Rawat for resistivity, R. J. Choudhary and M. Tripathi for
S-VSM, A. Wadikar for XPS and R. Sah for XAS measurements. We are
grateful to A. K. Sinha for support and encouragements.

\section*{References}
%\bibliography{TMN}


\end{document}